\documentclass[pra,aps,superscriptaddress,floats,showpacs,10pt]{revtex4-1}
\usepackage{graphicx,epsfig}
\usepackage{dcolumn}
\usepackage{bm}
\usepackage{amssymb}
\usepackage{amsmath}
\usepackage{textcomp}
\usepackage{hhline}
\usepackage{color}

\begin{document}

\title{Radiation Pressure Acceleration: the factors limiting maximum attainable ion energy}

\author{S. S. Bulanov}
\affiliation{Lawrence Berkeley National Laboratory, Berkeley, California 94720, USA}

\author{E. Esarey}
\affiliation{Lawrence Berkeley National Laboratory, Berkeley, California 94720, USA}

\author{C. B. Schroeder}
\affiliation{Lawrence Berkeley National Laboratory, Berkeley, California 94720, USA}

\author{S. V. Bulanov}
\affiliation{QuBS, Japan Atomic Energy Agency, Kizugawa, Kyoto, 619-0215, Japan}
\affiliation{A. M. Prokhorov Institute of General Physics RAS, Moscow, 119991, Russia}

\author{T. Zh. Esirkepov}
\affiliation{QuBS, Japan Atomic Energy Agency, Kizugawa, Kyoto, 619-0215, Japan}

\author{M. Kando}
\affiliation{QuBS, Japan Atomic Energy Agency, Kizugawa, Kyoto, 619-0215, Japan}

\author{F. Pegoraro}
\affiliation{Physics Department, University of Pisa and Istituto Nazionale di Ottica, CNR, Pisa 56127, Italy}

\author{W. P. Leemans}
\affiliation{Lawrence Berkeley National Laboratory, Berkeley, California 94720, USA}
\affiliation{Physics Department, University of California, Berkeley, California 94720, USA}

\begin{abstract}
Radiation pressure acceleration (RPA) is a highly efficient mechanism of laser-driven ion acceleration, with with near complete transfer of the
laser energy to the ions in the relativistic regime. However, there is a fundamental 
limit on the maximum attainable ion energy, which is determined by the group velocity of the laser. 
The tightly focused laser pulses have group velocities smaller than the vacuum light speed, and, 
since they offer the high intensity needed for the RPA regime, it is plausible that group velocity 
effects would manifest themselves in the experiments involving tightly focused pulses and thin foils. 
However, in this case, finite spot size effects are important, and another limiting factor, the transverse 
expansion of the target, may dominate over the group velocity effect. As the laser pulse 
diffracts after passing the focus, the target expands accordingly due to the transverse intensity profile of the laser. 
Due to this expansion, the areal density of the target decreases, making it transparent for radiation and effectively 
terminating the acceleration. The off-normal incidence of the laser on the target, due either to the experimental setup, 
or to the deformation of the target, will also lead to establishing a limit on maximum ion energy.                   
\end{abstract}

\pacs{52.25.Os, 52.38.Kd, 52.27.Ny} \keywords{ion accelerators, radiation pressure, relativistic plasmas} 
\maketitle

\section{Introduction}
{\noindent}The particle acceleration is one of the cornerstones of the fundamental physics, enabling 
different studies and applications. However, conventional technology of particle acceleration leads to large scale facilities, 
as well as high construction and operation costs. Hence there is a significant interest in advanced acceleration concepts that 
would reduce the size and cost of future accelerators. One of the most promising concepts is the laser plasma acceleration \cite{review_MTB, review_ESL}, 
where the particles are accelerated by strong electromagnetic (EM) fields generated by laser pulses in plasma. In particular, laser driven acceleration of electrons has made quite an impressive progress from 1979, when it was proposed \cite{Tajima&Dawson}, to 2014 when a 4.25 GeV beam 
was accelerated by a PW-class laser in the 9 cm plasma channel at the BELLA Center, LBNL \cite{4 GeV}.

The laser driven acceleration of ions \cite{review_MTB, review_ions1, review_ions2, review_ions3} has a longer history, starting from 
the scheme of acceleration proposed by Veksler in 1957 \cite{Veksler}. However this area of laser plasma acceleration only recently 
has become a very active area of research worldwide, due to numerous potential applications of laser driven ion sources, 
such as injectors for conventional accelerators \cite{injection}, hadron therapy of oncological diseases \cite{hadron therapy} 
(see for details review article  \cite{review_ions3} and references therein), 
radiography \cite{radiography}, nuclear physics studies \cite{nuclear}, studies of radiation damage and single 
event effect in electronics, 
as well as fast ignition inertial confinement fusion \cite{FI}, and drivers and probes for the studies of warm dense matter. 
This interest is also due to the recent availability of ultrahigh power 
lasers with focused intensity up to $10^{22}$ W/cm$^2$ \cite{10^22} and laser pulse cleaning techniques 
that allow a temporal intensity contrast of 14 orders of magnitude \cite{contrast}. New and efficient acceleration 
regimes were proposed, and some of them tested experimentally, giving rise to proton beams with the energy of about 
100 MeV from nm-scale foils of solid density \cite{70 MeV, BOA_exp, GIST_exp}.

The basic mechanisms of acceleration are (i) Target Normal Sheath Acceleration (TNSA) \cite{TNSA}, 
(ii) Coulomb Explosion (CE) \cite{CE}, (iii) Radiation Pressure Acceleration (RPA) \cite{RPA}, 
and (iv) Magnetic Vortex Acceleration (MVA) \cite{MVA1, MVA_new1, MVA_new2}. There are also several composite mechanisms, 
which are the either the combinations of basic ones or somehow the enhancement of the basic ones, such as Break-Out-Afterburner (BOA) \cite{BOA}, Shock Wave Acceleration (SWA) \cite{SWA}, Relativistic Transparency (RT) \cite{RT}, 
and Directed Coulomb Explosion (DCE) \cite{DCE}.  Also the use of composite targets (low density/high density 
or two layer - high Z/low Z) 
was proposed in a number of papers to either inject the ions into accelerating fields, enhance the interaction of 
the laser pulse with the high density part of the target, mitigate the effect of instabilities, or alter the accelerated 
ion spectra \cite{low+foil, two_stage, group velocity}. 

All these mechanisms can be parametrized in terms of  two dimensionless parameters, $\alpha_a$ and $\delta_a$, which are defined as
\begin{equation}
\alpha_a=\frac{a_0}{\varepsilon_p}\quad {\rm and} \quad \delta_a=\frac{d_e}{\lambdabar}\sqrt{a_0},
\label{eq:alpa-delpa}
\end{equation}
where we assume $a_0\gg1$. The first parameter $\alpha_a$  is the ratio of normalized laser EM field strength,
 $a_0=eE_0/m_e\omega c$ to normalized to $\varepsilon_p$ , which is a normalized surface density determined as \cite{Vshivkov}
\begin{equation}
\varepsilon_p= \frac{2 \pi e^2 n_e l}{m_e \omega c}.
\label{eq:epsilon-p}
\end{equation}
The second  parameter  $\delta_a$ is equal to the  collisionless skin-depth $\approx d_e\sqrt{a_0}$, 
with $d_e=c/\omega_{pe}$ and $\omega_{pe}=\sqrt{4\pi n_e e^2/m_e}$,   normalized on $\lambdabar=c/\omega$.
 We take also into account the relativistic dependence of the plasma frequency $\omega_{pe}/\sqrt{a_0}$ on the laser pulse amplitude assuming $a_0\gg 1$. 
Here $e$ and $m_e$ are the electron charge and mass,  $E_0$ and $\omega$ are the field strength and frequency, $c$ is the speed of light, $n_e$ is the electron density, and $l$ is the target thickness. The parameter $\alpha_a$ can be written in the form $\alpha_a=E_0/E_m$, where $E_m=2\pi e n_e l$ is the maximum electric field which can be produced by a charge slab of the density $n_e$ and thickness $l$.
The target can be considered as a thin foil provided its thickness $l$ is substantially less than the 
laser wavelength $\lambda=2\pi \lambdabar$. In Fig. \ref{fig-1} the dashed parabolic curve $l\approx \lambda$ subdivides the parameter plane in two sub-domains: a thin foil target for $\lambda>l$ and an extended plasma for $\lambda < l$.

The condition $a_0=\varepsilon_p$, i. e.  $\alpha_a=1$, marks the threshold transparency for laser pulses interactions with a thin foil target $l\ll \lambda$ (see Refs. \cite{Vshivkov} and \cite{LABASTRO-2015}). 
In Fig. \ref{fig-1}, the domains ($\rm{\bf I}$) and ($\rm{\bf II}$) correspond to overdense 
opaque and transparent slab, respectively. 
If the laser radiation interacts with an opaque target in the parameter domain ($\rm{\bf I}$), a
relatively small portion of hot electrons can leave the target, forming
a sheath with an electric charge separation. The corresponding
electric field accelerates ions in the TNSA regime \cite{TNSA}.
Above the line $a_0/\varepsilon_p=1$ ($\alpha_a=1$), i.e. in the domain ($\rm{\bf II}$), the laser radiation is so intense that it
blows out almost all electrons from the irradiated region of
the target. The remaining ions undergo fast expansion,
known as Coulomb explosion, due to the repelling of noncompensated
positive electric charges \cite{CE}. At the opaqueness/transparency
threshold, in the vicinity of the  $a_0/\varepsilon_p=1$ line in Fig. \ref{fig-1},
the conditions for ion acceleration in the RPA regime are realized \cite{RPA}.

In the regions  ($\rm{\bf III}$) and ($\rm{\bf IV}$), in Fig. \ref{fig-1}, where $d_e  \sqrt{a_0}/\lambdabar>1$, 
i. e. $\delta_a>1$, the plasma is underdense. We consider here the case of near-critical density plasma when the high power laser pulse undergoes the relativistic self-focusing \cite{LITVAK, SUN}. The laser pulse looses its energy with the energy depletion length equal to $l_{dep}=a_0 (\omega/\omega_{pe})^2 l_{las}$ (e.g., see Ref. \cite{MVA_new1}). For an ultra-short (several wavelength) laser pulse, $l_{las}\approx \lambda$, the choice of the plasma target thickness $l$ equal to the energy depletion length, $l=l_{dep}$, provides for maximum laser-target coupling \cite{DEPL}. 
This condition takes the form $l\approx a_0 (c \omega/\omega_{pe}^2)$, i.e., equivalent to the condition $a_0/\varepsilon_p\approx 1$ 
at the border between the domains ($\rm{\bf III}$) and ($\rm{\bf IV}$) in Fig. \ref{fig-1}. 
Under this condition the laser pulse deposits almost all 
its energy in the plasma producing a large number of fast electrons in the vicinity 
of the target rear surface. This results in  quasi-static magnetic field 
generation at the plasma vacuum interface by the electric current carried by fast electrons 
\cite{B-field-2}, which is crucially important for the magnetic vortex acceleration 
mechanism realization \cite{MVA1, MVA_new1, MVA_new2}. We note that effectively the ion acceleration mechanisms in Fig. 1 are bound by two conditions $d_e/\lambdabar\leq 1$ and $a_0\leq \varepsilon_{rad}^{-1/3}$, where $\varepsilon_{rad}$ is the parameter governing the strength of the radiation reaction effects \cite{NIMA}. The first of these boundaries states that for increasingly underdense plasma the ion acceleration mechanisms, which are mentioned in Fig. 1, cease to work. Such boundary for the MVA mechanism is mentioned in Ref. \cite{MVA_new1}. The second boundary corresponds to the threshold \cite{NIMA} of the radiation reaction and quantum recoil effects. These effects for laser intensities larger than $10^{23}$ W/cm$^2$ significantly alter the process of ion acceleration, which was illustrated in Refs. \cite{QED_ions} for RPA regime and in Ref. \cite{QED_transparency} for relativistic transparency.

Most of the laser ion acceleration experimental results were obtained in the TNSA regime \cite{review_ions1} with the maximum proton energy around 
70 MeV \cite{70 MeV}. Using a high contrast, 200 TW, femtosecond-pulse  laser irradiating a micron-thickness 
Al foil target, the TNSA regime has produced 40 MeV protons \cite{Ogura-2012}. 
Recently, several papers have been published that claim the experimental observation of the onset 
of the RPA regime of laser ion acceleration \cite{RPA_exp}, with the maximum energy up to 93 MeV \cite{GIST_exp}. 
In the case of long pulses, the acceleration of helium atoms up to 40 MeV from underdense plasma was observed at the 
VULCAN laser \cite{MVA_exp1}, the acceleration of protons up to 50 MeV at Omega EP \cite{MVA_exp2}, and there are some 
indications of reaching more than 100 MeV protons using the Trident laser \cite{BOA_exp}. Also, experiments with short 
pulses and cluster jets show that 10-20 MeV per nucleon ions can be generated via such interaction \cite{MVA_exp3}. 
The experimental data is summarized in Fig. \ref{fig-2}, along with the simulation results for proton acceleration 
with 1-3 PW and 10 PW laser pulses. 
We also show by area colored in gray the interval of proton energies relevant for the medical applications, 
which indicates that the experimentally 
obtained proton energies are at the threshold of being applicable for hadron therapy. 
This plot is an updated version of the one published in Ref. \cite{Fuchs}. 

\begin{figure}[h!]
\epsfxsize12cm\epsffile{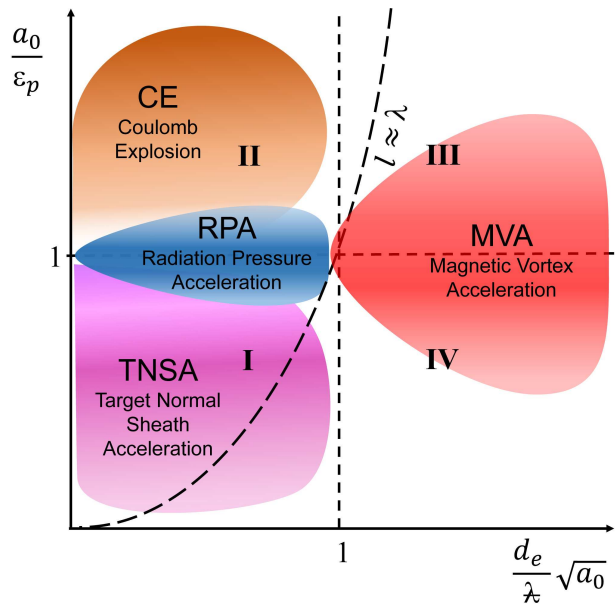}
\caption{\label{fig-1} The basic laser ion acceleration mechanisms in the plane of dimensionless parameters, $a_0/\epsilon_p$ and $d_e \sqrt{a_0}/\lambdabar$, characterizing the  laser amplitude and the target transparency.}
\label{fig-1}
\end{figure}

Each of these ion acceleration mechanisms has its own scaling of maximum ion energy 
with laser pulse power, intensity, or fluence. Each mechanism is characterized 
by the operational parameter range and requires specially 
designed targets in order to maximize the advantages of the particular mechanism 
and compensate for different limitations. For TNSA special target designs were considered, such as pizza-cone targets \cite{70 MeV}, 
nano structured  targets \cite{TNSA_nano targets}, thin foil targets with a thin 
layer deposited on the back, and thin foils with a low density slabs attached 
at the front \cite{low+foil,group velocity}. Recently several target designs 
were proposed for other mechanisms, including mass limited targets, RPA in a tube, 
and double shock formation in a gas jet for MVA \cite{MVA_double shock}.

\begin{figure}[h!]
\centering
\epsfxsize12cm\epsffile{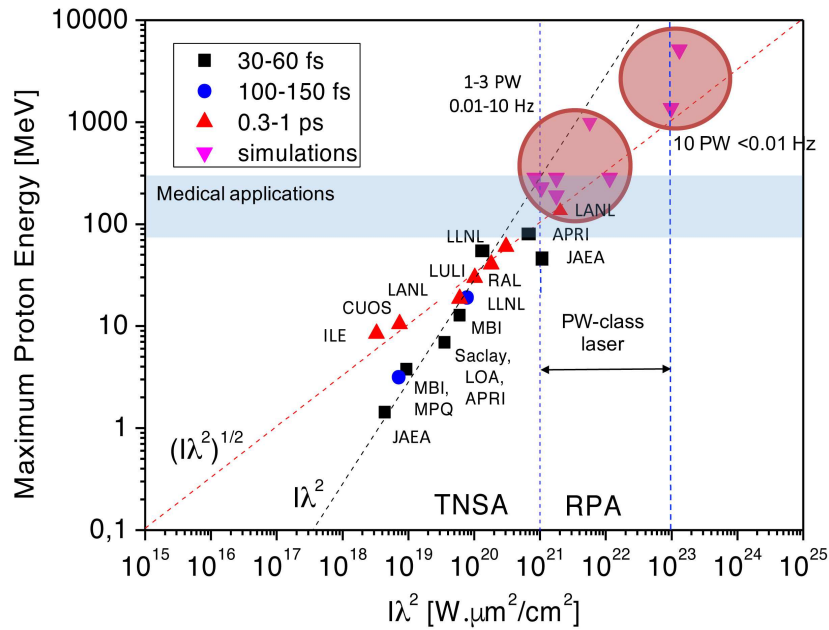}
\caption{\label{fig-2} The experimental data on maximum observed proton energies 
for different laser systems along with the simulation results for 1-3 PW laser pulses and 
10 PW laser pulses \cite{review_ions1, GIST_exp, BOA_exp}. The range of proton energies relevant 
for medical applications is shown by grey area.}
\label{fig-2}
\end{figure} 

In this work, we will address the RPA mechanism of ion acceleration and the factors limiting the maximum ion energy gain. 
RPA comes into play when the laser is able to push the foil as a whole by its radiation pressure. 
The idea goes back to the papers by Lebedev, Eddington \cite{RPA old}, and Veksler \cite{Veksler}, and 
has a close analogy, which was emphasized in many papers on the subject, to the ``light sail'' 
scheme for spacecraft propulsion \cite{RPA light sail}. The RPA is the realization of the relativistic 
receding mirror concept \cite{RM}. The role of a mirror is played by an ultra-thin solid density 
foil or by plasma density modulations emerging when the laser interacts with an extended under-critical 
density target, the so-called hole-boring RPA \cite{RPA_HB}.  
The problem of a plane EM wave reflection by a mirror moving with a relativistic velocity was considered 
by A. Einstein as an illustration of the Theory of Special Relativity \cite{Einstein}. 
The frequency of the reflected radiation is shifted down by a factor of $4\gamma_M^2$, 
where $\gamma_M$ is the Lorentz factor of the mirror. Thus the energy transferred 
to the mirror is $(1-1/4\gamma_M^2){\cal E}_{las}$, where ${\cal E}_{las}$  
is the energy of the laser pulse. For $\gamma_M\gg 1$ almost all laser energy is transferred to the foil, 
which makes this scheme very attractive in the ultrarelativistic limit \cite{RPA}. 
However there are a number of effects that limit the energy transfer from the laser to the ions. 
We aim at identifying these effects and finding means to either compensate for the limitations 
or completely remove them through target design, laser pulse shaping, or by changing 
the laser-target interaction. These effects are (i) target transparency \cite{optimized RPA, Macchi}, 
(ii) sub-luminal laser group velocity \cite{RPA_slow wave, group velocity}, 
(iii) transverse target expansion \cite{group velocity, Dollar}, and (iv) laser off-normal incidence. 
   
(i) Target transparency. In order to achieve high efficiency of energy transfer from the laser 
to the foil in the case of the RPA the foil should remain opaque for radiation during the acceleration process. 
However, opaque foils should either have high density or be rather thick, or both. This would increase the number 
of ions in the irradiated spot, thus decreasing the energy that the laser can transfer per ion. 
Therefore, if one wants to maximize the energy per ion then the acceleration should happen 
at the threshold of the foil transparency/opacity, which is governed by the reflection coefficient. 
At this threshold, the foil is opaque for radiation, but this opaqueness is ensured by the minimum 
possible number of ions. Thus, it was found in Ref. \cite{optimized RPA} the condition, 
$a(t)=\gamma(t)\varepsilon_p(t)$, must be realized to ensure optimal acceleration. 
This leads to an idea of laser pulse tailoring, so that the laser pulse has a special 
shape that would ensure this condition is satisfied every instant of 
the laser pulse interaction with the target. The matching of the laser pulse profile 
leads to significant reduction of the acceleration time and thus acceleration distance. 
This is a critical parameter for the RPA scheme of laser ion acceleration, 
since the acceleration distance is of the order of the required Rayleigh length 
for the high intensity laser systems. Therefore, the incident laser pulse of 
the form described in \cite{optimized RPA} offers a reasonable approach to compact 
laser ion accelerator, with relaxed requirements on the total laser pulse energy 
needed to achieve certain accelerated ion energy. 
   
(ii) Laser group velocity. Usually the energy gain in the RPA regime for the ultra-relativistic 
case is estimated using the relativistic mirror concept as $(1-1/4\gamma_M^2){\cal E}_{las}$. 
The effect of the EM wave group velocity being smaller than the vacuum light speed 
are not taken into account in this case. It is well known that group velocity effects play 
a major role in laser driven electron acceleration \cite{review_ESL} and should modify the RPA.  
If this effect is taken into account, then the energy gain in this case is proportional 
to the difference between the instantaneous foil velocity, 
$\beta$, and the laser group velocity, 
$\beta_g$, $\Delta {\cal E}\approx 2\gamma^2\beta(\beta-\beta_g){\cal E}_{las}$ 
\cite{RPA_slow wave, group velocity}. A laser pulse can not accelerate the foil to the velocity larger than its group velocity. 
   
(iii) Transverse target expansion. The high intensity needed for the RPA regime 
is usually offered by tightly focused laser pulses, which have group velocities 
smaller than the vacuum light speed. So it is plausible to expect that the group velocity 
effects will manifest themselves in  experiments involving such tightly focused pulses and thin foils. 
However, in this case, finite size spot effects \cite{Dollar} are important, and another 
limiting factor, the transverse target expansion will dominate in limiting maximum 
ion energy \cite{group velocity}. The transverse expansion is caused by the diffraction 
of the tightly focused laser pulse, resulting in the decrease of the target areal density, 
which effectively terminates the acceleration. 
   
The utilization of external guiding may relax the constraints 
on maximum attainable ion energy mainly imposed by transverse expansion. 
Namely, the use of a composite target having a thin foil followed by an NCD slab. 
The NCD slab provided guiding of the laser pulse during the acceleration process. 
The comparison of a single foil RPA and a composite target RPA shows that, in the latter case, 
the ions have energy several times larger than in the former case, thus greatly increasing 
the effectiveness of the RPA regime of laser-driven ion acceleration. 
In such a configuration, the group velocity effects begin 
to dominate and determine the maximum achievable ion energy \cite{group velocity}.   
     
(iv) Laser off-normal incidence. The laser ion acceleration by the radiation pressure 
in the case of an off-normal incidence of the laser pulse gives rise to another 
limit to the maximum attainable ion energy. Off-normal incidence is routinely 
used in  many experiments on laser ion acceleration to reduce the chance 
of damaging the laser system by the reflected light. The limit on the maximum 
ion energy is determined by the angle of incidence. If the off-normal incidence 
is considered for the pulse with $\beta_g<1$, then the maximum ion velocity is $\beta_{max}=\beta_g\cos\theta$. 
The limit is due to the fact that at some moment during the interaction 
the foil velocity becomes so large that the longitudinal component 
of the EM wave vector vanishes in the rest frame of the foil. 
This effect implies that the tightly focused pulses are not beneficial 
for the RPA of ions. For such pulses, the foil quickly becomes deformed during the 
initial stage of interaction, resulting in a curved target. 
Then, locally the interaction can be viewed as an off-normal incidence 
with the limiting velocity $\beta_g\cos \theta(r,z)$, where  $\theta(r,z)$ 
corresponds to the local value of the angle of incidence. This results in a power-law spectrum of ions, 
preventing the development of the mono energetic features.
     
The paper is organized as follows. In section 2 we review the effects of the target transparency 
on the RPA regime and the laser pulse shaping to compensate for these effects \cite{optimized RPA}. 
In section 3 we review the group velocity limitations \cite{RPA_slow wave,group velocity}. 
The transverse expansion effects  \cite{group velocity} are reviewed in section 4. 
The consequences of the off-normal incidence of a laser  pulse on a foil are addressed in section 5, 
together with the preliminary considerations on the interaction of the laser pulse 
with a deformed target in the RPA regime. We conclude in section 6. 
     
\section{Target transparency effects}

In what follows we review the effects of target transparency on the RPA regime. 
It has been noted in a number of papers \cite{optimized RPA, Macchi} that evolving target transparency 
significantly changes the evolution of the ion energy. The target transparency leads 
to reduction in acceleration effectiveness and to the notion of the optimal target thickness 
that maximizes the ion energy for fixed laser pulse. The motion 
of the foil under the action of the laser pulse radiation pressure is described by the equation 
(we set $c=1$ below throughout the paper) \cite{RPA}
\begin{equation} 
\frac{1}{\left(1-\beta^2\right)^{3/2}}\frac{d\beta}{dt}
=\frac{(2|\rho|^2+|\alpha|^2)\left|E_L(\psi)\right|^2}{4\pi n_e l}\left(\frac{1-\beta}{1+\beta}\right), \label{eqn1}
\end{equation}
where $\beta=dx/dt$ is the foil velocity, $x(t)$ is the position of the foil, and 
$E_L(\psi)$ is the laser pulse field, which depends on the variable $\psi=t-x(t)$.  
Here $\rho$ is the reflection coefficient and $\alpha$ is the absorption coefficient. 
These coefficients are connected through the energy conservation condition, $|\rho|^2+|\tau|^2+|\alpha|^2=1$, 
where $\tau$ is the transmission coefficient. The solution of Eq. (\ref{eqn1}) for the foil initially at rest, 
$\beta(0)=0$,  assuming total reflection (i.e., $\rho=1$), is 
\begin{equation}
\beta=\frac{W(2+W)}{2+2W+W^2}~~~\text{or}~~~\gamma=\frac{2+2W+W^2}{2(1+W)}.
\end{equation} 
Here $W=2\mathcal{F}_L/n_e l$ is the normalized is the laser pulse fluence (incident laser energy per unit area),
\begin{equation}
\mathcal{F}_L=\int\limits_0^\infty\frac{|E_L(\psi)|^2}{4\pi} d\psi
\end{equation}
In the ultra-relativistic case the energy of the foil asymptotes to  \cite{RPA}
\begin{equation}
\gamma^{max}=\frac{\mathcal{F}_L}{n_e l},
\end{equation}
which illustrates the fact that in the ultra-relativistic case almost 
all laser energy is transferred to the foil energy ${\cal E}=m_p c^2 \gamma^{max}$ \cite{RPA}.

The equation of motion (\ref{eqn1}) can be rewritten in terms of dimensionless variables:
\begin{equation}
\frac{1}{\left(1-\beta^2\right)^{3/2}}\frac{d\beta}{dt}=\kappa\frac{1-\beta}{1+\beta},~~~\text{where}~~~\kappa=\frac{(2|\rho|^2+|\alpha|^2)\left|E_L(\psi)\right|^2}{4\pi n_e l\omega}=\frac{(2|\rho|^2+|\alpha|^2)}{2}\frac{m_e}{m_i}\frac{a^2(\psi)}{\varepsilon_p}
\end{equation}
with the dimensionless parameter $\varepsilon_p$ defined by Eq. (\ref{eq:epsilon-p}).
Time is measured in units of $\omega^{-1}$.

The fact that the foil may be transparent for radiation significantly modifies the acceleration process. Consider the amplitude of the EM wave reflected by a thin ($l\ll\lambda$) foil, \cite{Vshivkov, optimized RPA, LABASTRO-2015}  
\begin{equation}\label{a_r}
a_r=\varepsilon_p\sqrt{\frac{\sqrt{(a_0^2-\varepsilon_p^2-1)^2+4a_0^2}
+a_0^2-\varepsilon_p^2-1}{\sqrt{(a_0^2-\varepsilon_p^2-1)^2+4a_0^2}+a_0^2-\varepsilon_p^2+1}}.
\end{equation}
In the two limiting cases of opaque ($a_0\ll \varepsilon_p$) and transparent ($a_0\gg\varepsilon_p$) 
foil, i.e. in the domains ($\rm{\bf I}$) and  ($\rm{\bf II}$) shown in Fig. 1, respectively, we find
\begin{equation}
a_r=\left\{
\begin{tabular}{l}
$\displaystyle\frac{\varepsilon_pa_0}{\sqrt{1+\varepsilon_p^2}},~~~a_0\ll min\left[1,\varepsilon_p\right] $\\ \\
$\displaystyle\varepsilon_p-\frac{\varepsilon_p}{2a_0^2},~~~a_0\gg max\left[1,\varepsilon_p\right].$
\end{tabular}
\right.
\end{equation} 
In the case of an opaque foil almost all radiation is reflected and $a_r$ 
is determined by the incident EM field amplitude. In the case of a transparent foil, 
the reflected EM wave amplitude is determined by the current 
the incident wave can drive in the foil, which is proportional to $\varepsilon_p$. 

\begin{figure*}[tbp]
\epsfxsize5cm\epsffile{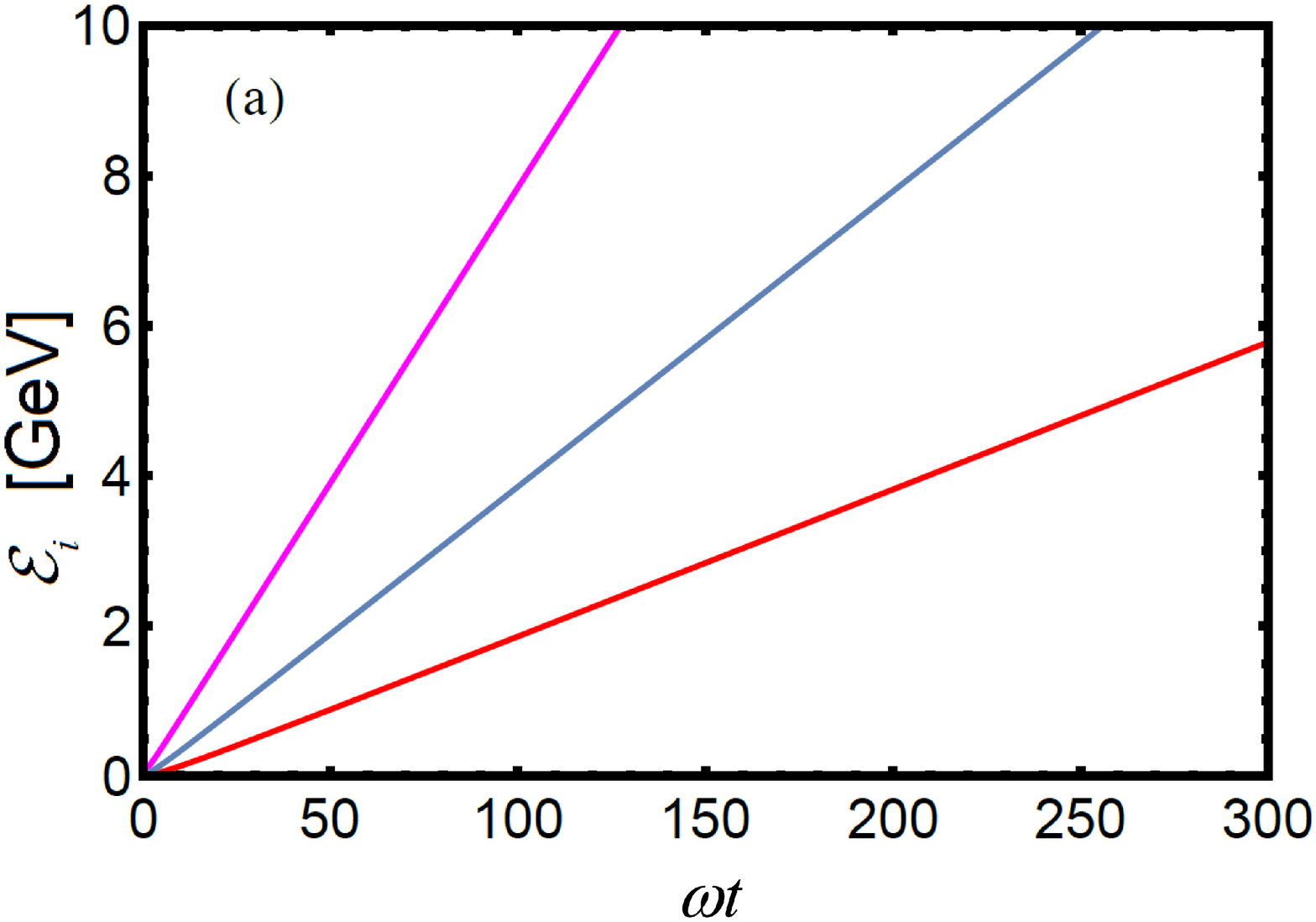}
\epsfxsize5cm\epsffile{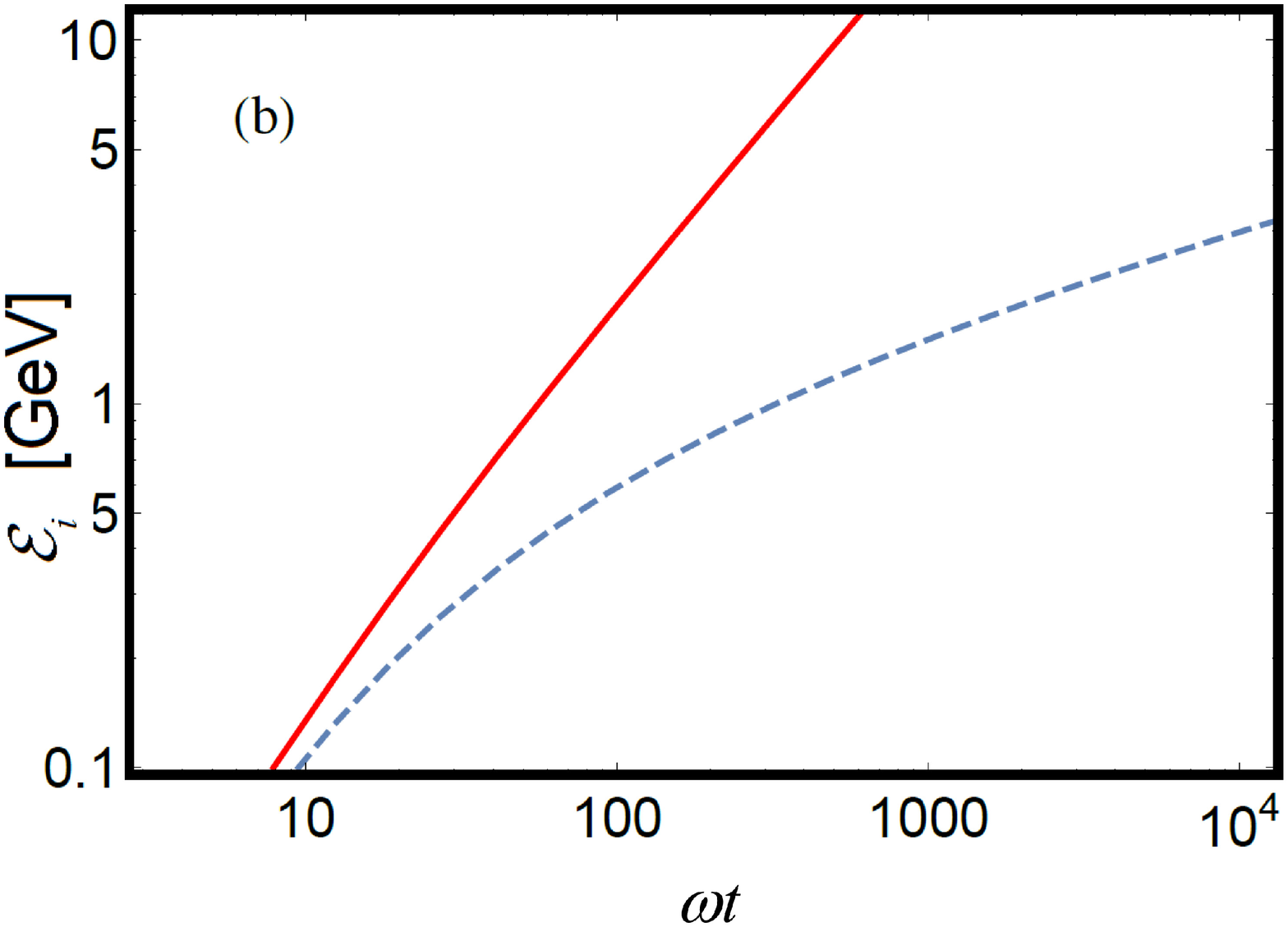}
\epsfxsize5cm\epsffile{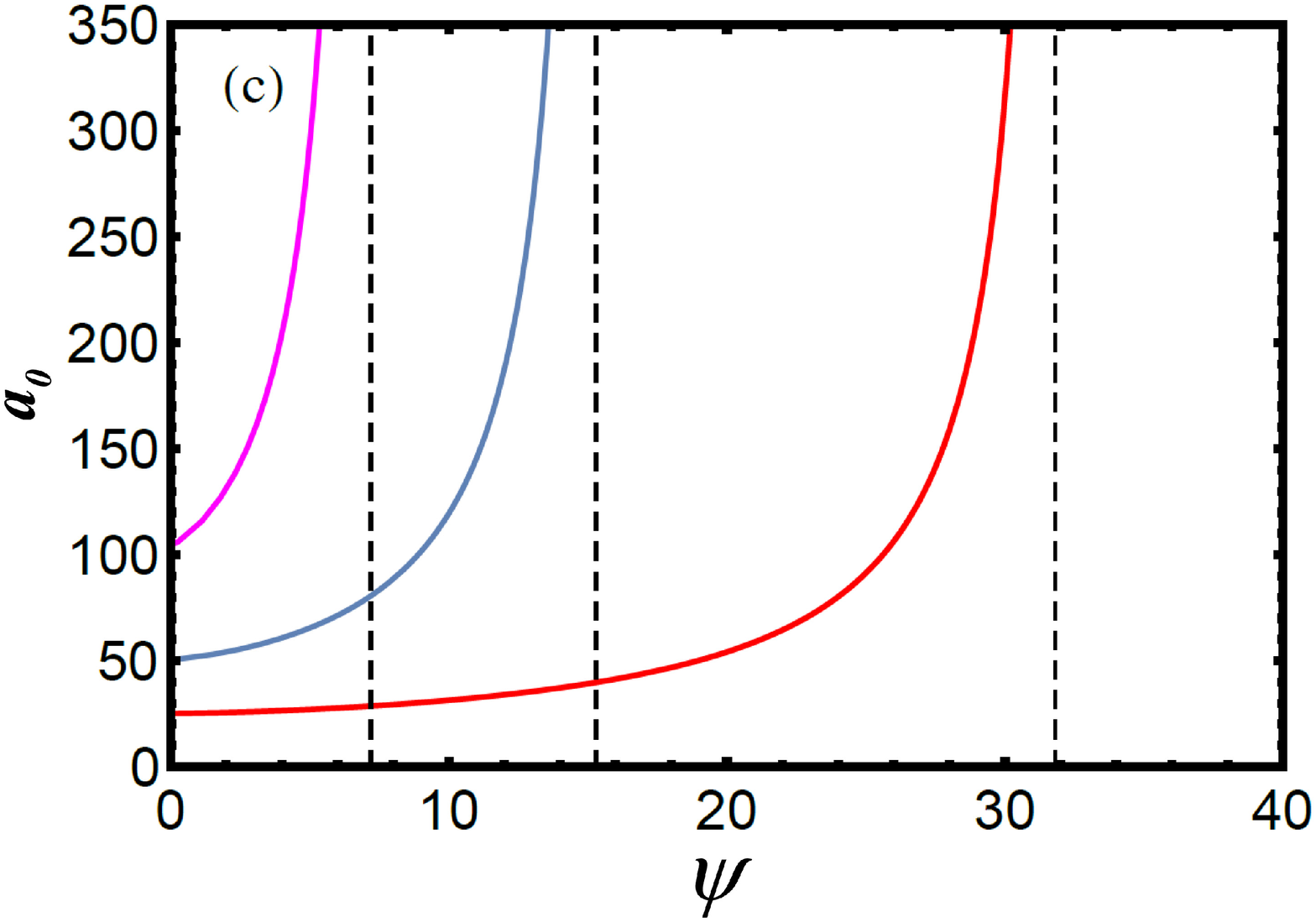}
\caption{(a) The evolution of the maximum ion energy for different values 
of the normalized target density  $\varepsilon_p$: $\varepsilon_p=25$ (red curve), 
50 (blue curve), and 100 (magenta curve); 
(b) the comparison of the evolutions of the maximum ion energy of 
a profiled laser pulse (solid line) and a constant amplitude pulse 
(dashed line) for $\varepsilon_p=25$; (c) the profiles of the laser pulses 
for $\varepsilon_p=25$ (red curve), 50 (blue curve), and 100 (magenta curve), 
which are able to maintain the condition $a_0=\gamma\varepsilon_p$ during the acceleration process. 
Dotted vertical lines denote the position of $\psi_*$ 
for each of the three values of $\varepsilon_p$: $\psi_*(\varepsilon_p=25)=31.8$, 
$\psi_*(\varepsilon_p=50)=15.3$, and $\psi_*(\varepsilon_p=100)=7.2$.}
\label{fig-3}
\end{figure*}

The parameter $\varepsilon_p$ plays an important role in determining the optimal regime of acceleration (see Fig. \ref{fig-1}) \cite{optimal}. 
For a foil at rest the condition $a_0=\varepsilon_p$ marks the threshold of 
opacity/transparency of the foil to radiation \cite{Vshivkov}. 
At this threshold the opacity is maintained by the minimum possible number of ions, 
which means that if the EM wave transfers some amount of its energy to the foil, 
the amount of absorbed energy per ion is maximal. For the case of non-relativistic 
ion energies the condition $a_0=\varepsilon_p$ marks the regime of laser-foil interaction, 
which maximizes the accelerated ion energy (see Ref. \cite{optimal}). However, in the general case, 
the foil velocity, as well as the fact that $\varepsilon_p$ is not relativistically invariant, should be taken into account. 
Therefore the condition $a_0=\varepsilon_p$ should be replaced by $a_0=\gamma\varepsilon_p$ \cite{optimized RPA}. 
In order to maximize the effect of the matching of the laser pulse to the foil the condition  
$a_0=\gamma\varepsilon_p$ should be satisfied at every instant of time during the acceleration process. 
To achieve such matching a tailored laser profile is required. Assuming that this condition is satisfied, the equation of the foil motion 
is
\begin{equation}\label{gamma_epsilon=a0}
\frac{d\beta}{dt}=\varepsilon_p\frac{m_e}{m_i}\rho^2\frac{(1-\beta)^{3/2}}{(1+\beta)^{1/2}}.
\end{equation} 
This equation can be solved numerically, and the results are shown 
in Fig. \ref{fig-3} for $\varepsilon_p=25,~50,~\text{and}~100$. 
Here we assumed that the foil is made of hydrogen,  $m_i=m_p$. 
The dependence of the proton energy on time demonstrates linear behavior (Fig. \ref{fig-3}a). 
This is drastically different from the case of a constant amplitude laser pulse, 
where the energy grows as $\sim t^{1/3}$ \cite{RPA} (see Fig. \ref{fig-3}b). 
Therefore in the case of a profiled laser pulse the time needed 
to reach certain ion energy is significantly smaller, which may lead to a reduction in size of laser ion accelerators.        

The condition, $a_0=\gamma\varepsilon_p$, which was used in deriving Eq. (\ref{gamma_epsilon=a0}), 
allows reconstruction of the profile of the laser pulse, which is shown in Fig. \ref{fig-3}c. 
The laser field determined by this profile is inversely proportional to $\psi-\psi_*$, where $\psi_*$ 
is a function of the parameter $\varepsilon_p$. In order to study the behavior 
of the profiled laser pulse analytically and estimate the value $\psi_*$, the case of ultrarelativistic ion energies is considered. 
For the ultra-relativistic case, the field of the laser pulse can be found from the solution of
 Eq. (\ref{gamma_epsilon=a0}) \cite{optimized RPA}:
\begin{equation}\label{field}
a=\frac{\varepsilon_p}{1-\psi/\psi_*},~~~\text{where}~~~\psi_*=\frac{m_i}{m_e}\frac{2}{\varepsilon_p} 
\end{equation}
is the maximum duration of the pulse. If we compare this solution, Eq. (\ref{field}), 
with the numerical solution of Eq. (\ref{gamma_epsilon=a0}) (see Fig. \ref{fig-3}), 
then, first, one can see the inverse proportionality of the field amplitude to $\psi-\psi_*$, 
and, second, the analytical estimate for $\psi_*$ approximately reproduces the numerical solution 
(see Fig. \ref{fig-3}c). Note here that one of the most serious 
issues with the RPA of thin foils is the development of the Rayleigh-Taylor (RT) 
instability \cite{RT instability} (see also Ref. \cite{COMREN}), whose growth rate is comparable 
with the acceleration time, and which can effectively terminate the acceleration. 
However, according to the results of Ref. \cite{unlimited, RT instability}, 
the laser profile given by Eq. (\ref{field}) suppresses the development of the RT instability.  

The numerical solution of the equation 
of motion indicates that the profiled laser pulse is able to accelerate ions 
to some fixed energy in a significantly smaller period of time, 
than a constant amplitude laser pulse. Using the solutions of the equation of motion 
in these two cases yields the relation between the acceleration times assuming 
that both pulses have the same total energy. The durations, 
$\tau_{las}$ and $T_{las}$, are related to each other by $T_{las}=\tau_{las}/(1-\tau_{las}/\psi_{*} )$, where $T_{las}$ is for a constant amplitude pulse and $\tau_{las}$ for profiled laser. In the limit of $t\rightarrow\infty$ the acceleration times are \cite{optimized RPA}
\begin{equation}
T_{acc}\approx \frac{1}{6}W_\rho^2T_{las}~~~\text{and}~~~
t_{acc}=\frac{W_\rho^2}{2}\frac{\psi_*^2}{\tau_{las}^2}\left(1-\frac{\tau_{las}}{\psi_*}\right)\tau_{las},
\end{equation}
for the constant amplitude and profiled laser, respectively. The acceleration time in the case of a constant amplitude laser pulse 
is significantly longer than in the case of a profiled pulse:
\begin{equation}
\frac{T_{acc}}{t_{acc}}=\frac{1}{3}\left(\frac{\tau_{las}/\psi_*}{1-\tau_{las}/\psi_*}\right)^2.
\end{equation}
For example, let us assume that the profiled pulse duration is $\tau_{las}=0.9\psi_*$, 
which is ten times smaller then the duration of the constant amplitude pulse having 
the same total energy. The acceleration time for the profiled pulse in this case 
is approximately one hundred times smaller: $t_{acc}\approx 10^{-2}T_{acc}$. 
Thus the profiling of laser pulses may lead to a significant reduction of the 
length of laser ion accelerators based on the RPA mechanism. 

\section{Group velocity effect}

{\noindent} The group velocity of the laser pulse being smaller 
than the vacuum light speed imposes a fundamental limit on the maximum attainable ion energy during the RPA 
of a thin foil. The naive estimate shows that the target energy gain is proportional to the difference between 
the instantaneous foil velocity and the laser group velocity \cite{RPA_slow wave, group velocity},
\begin{equation} 
\Delta {\cal E}\approx 2\gamma^2\beta(\beta-\beta_g){\cal E}_{las}.
\end{equation} 
Thus, in principle, the laser pulse with $\beta_g<1$ can only accelerate the foil up to the velocity 
equal to its group velocity $\beta^{max}=\beta_g$. If we take this effect into account, then the equation 
of motion of the foil under the action of the laser pulse radiation pressure takes the form different 
from the one used in the previous section \cite{RPA_slow wave,group velocity}:
\begin{equation}
\label{eq of motion}
\frac{d\beta}{dt}=\kappa\beta_g(1-\beta^2)^{1/2}(\beta_g-\beta)(1-\beta\beta_g).
\end{equation}  
From Eq. (\ref{eq of motion}) one can see that if the instantaneous velocity of the foil becomes 
equal to the laser group velocity, the right hand side of the equation vanishes, the acceleration is terminated, and the foil has reached maximum attainable velocity. The solution of Eq. (\ref{eq of motion}) can be written in the following form for $\beta(0)=0$:
\[\left\{\ln\frac{(1-\beta\beta_g+(1-\beta_g^2)^{1/2}(1-\beta^2)^{1/2}\beta_g}{(\beta_g-\beta)(1+(1-\beta_g^2)^{1/2})}\right .\]
\begin{equation}
\left . -\beta_g\left[\arctan\frac{(1-\beta_g^2)^{1/2}(1-\beta^2)^{1/2}}{\beta_g-\beta} 
-\arccos \beta_g\right]\right\} =\beta_g(1-\beta_g^2)^{3/2}K_\beta(t),
 \label{eq of motion_time}
\end{equation}
%
where 
\begin{equation}
K_\beta(t)=\int\limits_0^t \kappa dt^\prime.
\end{equation} 
When time tends to infinity and assuming constant amplitude laser pulse, $K_\beta(t)=\kappa t$, the following scaling for the foil velocity is obtained as $\beta$ approaches $\beta_g$ (here it is assumed that the foil stays opaque during the acceleration process, $|\rho|=1$):
\begin{equation}\label{group velocity limit}
\beta=\beta_g-\exp\left(-\beta_g(1-\beta_g^2)^{3/2}\kappa t\right). 
\end{equation}
Thus the ion energy is limited by $\gamma_g=1/\sqrt{1-\beta_g^2}$.

We note that the limitation of the achievable ion energy by  $\gamma_g$ implies the adiabatic regime 
of the laser pulse interaction with a thin foil target. In the case of non-adiabatic interaction (it may correspond to the laser pulse with 
a gradual profile, which requires additional consideration, which will be carried out in the forthcoming publications), the upper limit on the accelerated ion energy is imposed by $2 m_i c^2 \gamma_g^2$. We may compare the laser wake field acceleration scaling in 'non-adiabatic' \cite{Tajima&Dawson, review_ESL} and 'adiabatic' \cite{unlimLWFA} regimes. In the standard 'non-adiabatic', regime the electron energy is proportional to the square of the Lorentz gamma-factor calculated for the laser-driver group velocity, $\gamma_e\approx 2 \gamma_g^2$, \cite{Tajima&Dawson, review_ESL}. 
The 'adiabatic' case is realized for the laser-target parameters providing the non-phase-slippage conditions 
when $\gamma_e\approx 2 \gamma_g^2$ \cite{unlimLWFA}, which results in unlimited electron acceleration.

In Fig. 4a we present the numerical solution of Eq. (\ref{eq of motion_time}) 
for Gaussian pulses with a duration of 27 fs (10 cycles), $\lambda=800$ nm wavelength, 
f-number of $f/D=1$ and $f/D=1.5$, interacting with  a $0.25\lambda$ thick hydrogen foil 
with an electron density of $400n_{cr}$. The evolution of the maximum ion energy is shown for 
the averaged laser power of 1.8 PW. One can see that for $f/D=1.5$ 
the difference between the $\beta_g=1$ and $\beta_g<1$ curves is small since 
the laser fluence is not high enough to push the ions up to the energy $\gamma_g=1/\sqrt{1-\beta_g^2}$. 
However the difference is clearly seen in the case of $f/D=1$. The curve corresponding to $\beta_g<1$ 
is limited by the group velocity, while the $\beta_g=1$ curve continues to grow. 
Thus the numerical solution of the equation of motion demonstrates the effect 
of the laser group velocity being smaller than the vacuum light speed.

\begin{figure*}[tbp]
\epsfxsize15cm\epsffile{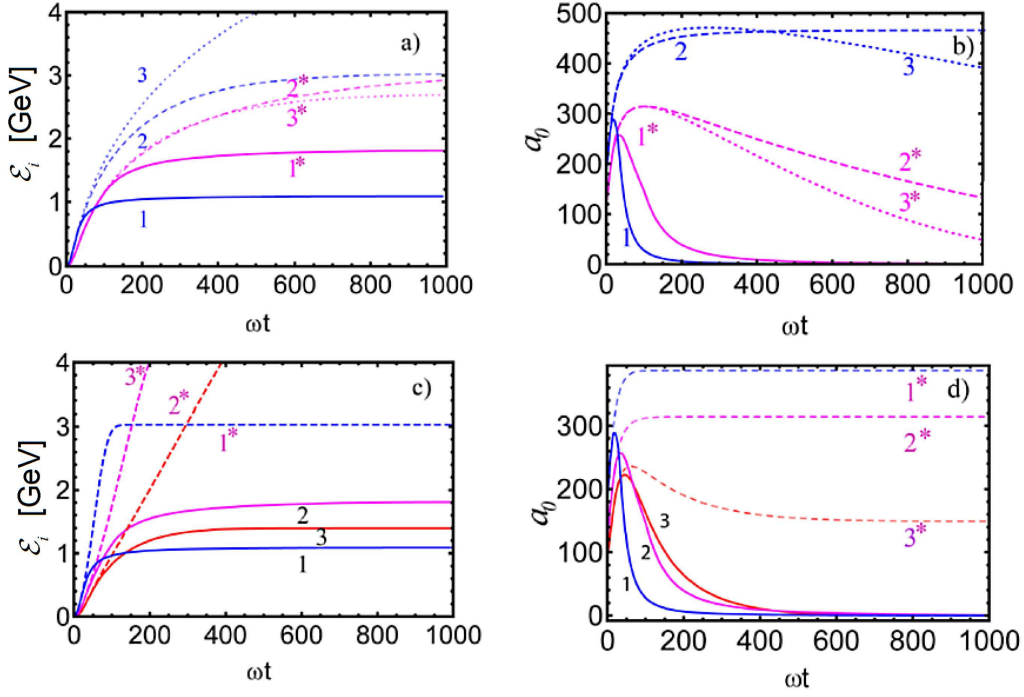}
\caption{(a) The evolution of the maximum ion energy for two values of f-number: 
f/D=1 (blue curves, 1, 2, and 3) and f/D=1.5 (magenta curves, 1$^*$, 2$^*$, and 3$^*$). 
The solid curves (1 and 1$^*$) are the solutions of Eq. (\ref{eq of motion_x}) 
with laser divergence and target expansion taken into account. 
The dotted curves (3 and 3$^*$) are the solutions of Eq. (\ref{eq of motion}) 
with $\beta_g=1$ and dashed curves (2 and 2$^*$) are the solutions of Eq. (\ref{eq of motion}) 
with $\beta_g<1$. 
(b) The evolution of EM field amplitude at the foil corresponding to the curves in Fig. 4a. 
(c) The evolution of the maximum ion energy for f/D=1 (blue curves, 1 and 1$^*$), 
f/D=1.5 (magenta curves, 2 and 2$^*$), and f/D=2 
(red curves, 3 and 3$^*$). The curves 1, 2, and 3 are the solutions 
of Eq. (\ref{eq of motion_x}), while the curves 1$^*$, 2$^*$, and 3$^*$ 
correspond to the guided case, where the laser is not diffracting 
and the reflection coefficient is equal to one. 
(d) The evolution of EM field amplitude at the foil corresponding to 
the curves in Fig. 4c. Laser pulse power is 1.8 PW, duration is 30 fs, 
the foil thickness is $0.25\lambda$, and density is $n_e=400 n_{cr}$.}
\label{fig-4}
\end{figure*}

\section{Transverse expansion effect}     

{\noindent} The finite focal spot effects 
play an important role in laser driven ion acceleration, especially using the RPA mechanism. 
These effects tend to cause the transverse expansion of the target, thus lowering target 
surface density $n l$ and increasing transparency. 
According to the results of the previous 
section, the increased target transparency reduces the effectiveness 
of acceleration and leads to the termination of the acceleration process. 
RPA usually requires high laser intensities to operate. 
High intensities are generated by tightly focussing the laser radiation to a small focal spot, 
which can be only a few wavelengths in size. Such a configuration causes laser group velocity 
to be noticeably  smaller than the vacuum light speed, $\beta_g\simeq 1-1/k^2w_0^2$ ($w_0$ 
is the laser pulse waist at focus) \cite{group velocity}. Also such laser pulses quickly 
diverge after passing the focus, with Raleigh length of several wavelengths, $L_R=\pi w_0^2/\lambda$, 
and the amplitude of the field decreasing with the distance from the focus as $a(x)=a_0/\sqrt{1+(x/L_R)^2}$. 
The simplest model for the transverse expansion is to assume that the expansion is due to the laser pulse divergence, and the surface 
density decrease is inversely proportional to the laser waist increase squared:
\begin{equation}\label{expansion}
w(x)=w_0\sqrt{1+(x/L_R)^2}~~~\text{and}~~~n_e l=n_0l_0\left[1+(x/L_R)^2\right]^{-1}.
\end{equation}       
In order to detrmine the effect of the transverse expansion on the maximum attainable ion energy, 
we consider the acceleration of an on-axis element of the foil. In this case 
both the laser field and the foil element can be approximated by an expanding spherical 
cup with curvature radius equal to the laser waist, $w(x)$. The transparency parameter  $\varepsilon_p$ then scales as 
\begin{equation}\label{expansion epsilon}
\varepsilon_p(x) = \varepsilon_p(0)\left[1+(x/L_R)^2\right]^{-1}
\end{equation}
with the distance from focus. Substituting Eqs. (\ref{expansion},\ref{expansion epsilon}) in the equation of motion (\ref{eq of motion}), we get the following equation for the expanding foil under the action of a diverging laser pulse: 
\begin{equation}\label{eq of motion_x}
\frac{d\beta}{dt}=|\rho(x)|^2\frac{m_e}{m_i}\frac{a^2(\psi)}{\varepsilon_p}\beta_g(1-\beta^2)^{1/2}(\beta_g-\beta)(1-\beta\beta_g).
\end{equation} 
Note that the right hand side of the Eq. (\ref{eq of motion}) depends on coordinate $x$ only through the reflection coefficient \cite{optimized RPA} (see section 2). Thus the main contribution of the transverse expansion is the increased target transparency as the acceleration evolves. It is plausible to assume that it will result in earlier termination of the acceleration and lower maximum attainable ion energy.   

The Eq. (\ref{eq of motion_x}) can be solved numerically, allowing for the comparison between the cases with and without the transverse expansion taken into account. The results of the solution are shown shown in Fig. 4 for a $1.8$ PW average power laser pulse interacting with a $0.25\lambda$ thick hydrogen foil with the density of $n_e=400 n_{cr}$ for three values of the f-number: $f/D=1$, $f/D=1.5$, and $f/D=2$. There are two sets of curves in Fig. 3a: (i) corresponding to f/D=1 and (ii) corresponding to f/D=1.5. In each set we compare three cases: no target expansion and laser divergence, and $\beta_g=1$, no target expansion and laser divergence, and $\beta_g<1$, and target expansion and $\beta_g<1$. The target expansion dominates in both sets, severely limiting the maximum attainable ion energy by early termination of the acceleration process. This can be illustrated by the evolution of the value of the laser field amplitude at the foil, see Fig. 3b. The transverse expansion results in lower maximum value of the field as well as the duration of its interaction with the foil.    

In order to compensate for the effects of the transverse expansion and relax the limits on maximum attainable ion energy, the laser foil interaction should be modified in a way that would preserve the surface density of the foil. This would result in the foil being opaque for radiation during the acceleration. We propose to employ an external guiding to satisfy the above mentioned requirements. The guiding can be modeled assuming that the laser pulse is guided in a self-generated channel and the foil stays opaque for the pulse. We solve the Eq. (\ref{eq of motion_x}) numerically for the three values of the guided laser pulse waist  $w_0=0.9\lambda$ ($f/D=1$), $w_0=1.35\lambda$ ($f/D=1.5$), and $w_0=1.8\lambda$ ($f/D=2$) (Fig. 3c). Such interaction configuration leads to a significant enhancement of the maximum attainable ion energy, which is now limited by the group velocity of the laser in the guiding structure instead of being limited by the increasing transparency. This enhancement can also be illustrated by the evolution of the laser field amplitude at the foil. In the guided case its maximum amplitude is not only higher than in the unguided case, but also the pulse becomes phase-locked with the foil, meaning that the maximum attainable ion energy is reached.   

As for the target design that would demonstrate the qualities described above, we propose to employ a composite target, consisting of a thin foil, followed by a near critical density (NCD) slab \cite{group velocity}. The laser pulse will generate a channel in the NCD plasma \cite{MVA1, MVA_new1, MVA_new2, helium} both in electron and ion densities. The laser will propagate in this channel pushing the irradiated part of the foil in front of it. Though the foil density will drop due to the transverse expansion, the NCD plasma electrons being snowplowed by the pulse would provide an opaque density spike, which being pushed by the radiation pressure would drag the ions of the foil with it. Thus such configuration is similar to the one considered above: the laser pulse is guided with no diffraction and, although the density of ion decreases, the reflection coefficient is unity. The results of the computer simulations of such target configuration, reported in Ref. \cite{group velocity}, show the significant enhancement of the maximum ion energy in the case of a composite target compared to the case of a single foil target.  
     
\section{Off-normal incidence}     

In this section we consider the acceleration of a thin solid density foil by the off-normal incidence laser pulse in the RPA regime. The laser wave vector has an angle, $\theta$, with target normal. Consider two Lorentz transformations to the reference frame where the foil is at rest and the wave is normally incident on the foil. The first transformation is a boost along the x-axis to the reference frame moving with the velocity $\beta$. The transformation matrix is
\begin{equation}
B_x(\beta)=
\left(
\begin{tabular}{cccc}
$\gamma$ & $-\beta\gamma$ & $0$ & $0$ \\
$-\beta\gamma$ & $\gamma$ & $0$ & $0$ \\
$0$ & $0$ & $1$ & $0$ \\
$0$ & $0$ & $0$ & $1$ 
\end{tabular}
\right).
\end{equation}     
The second transformation is a boost along the y-axis to eliminate the x component of the field:
\begin{equation}
B_y(\beta^\prime)=
\left(
\begin{tabular}{cccc}
$\gamma^\prime$  & $0$ & $-\beta^\prime\gamma^\prime$ & $0$ \\
$0$ & $1$ & $0$ & $0$ \\
$-\beta^\prime\gamma^\prime$ & $0$ & $\gamma^\prime$  & $0$ \\
$0$ & $0$ & $0$ & $1$ 
\end{tabular}
\right).
\end{equation}
After the first transformation the EM wave vector is
\begin{equation}
k_\mu^\prime=\left(\gamma\omega-k_x\beta\gamma,\gamma k_x -\omega\beta\gamma,k_y,0 \right),
\end{equation}
and the angle of the laser pulse incidence on the foil becomes
\begin{equation}
\cos\theta^\prime=\frac{\cos\theta-\beta}{1-\beta\cos\theta}.
\end{equation}  
Note that when $\beta=\cos\theta=k_x/\omega$ the x component of the wave vector vanishes, and, in such reference frame, the EM wave propagates parallel to the foil. After the second transformation the EM wave vector is
\[ k_\mu^{\prime\prime}= \]
\begin{equation}
\left\lbrace\gamma^\prime[-k_y\beta^\prime+\gamma\left(\omega-k_x\beta\right)],
\gamma [k_x-\omega\beta],\gamma^\prime k_y-\gamma\left(\omega-k_x\beta\right),0\right\rbrace.
\end{equation} 
The condition $k_y^{\prime\prime}=0$, \textit{i.e.}, the reference frame where the  EM wave is normally incident on the foil, yields
\begin{equation}\label{beta^prime}
\beta^\prime=-\frac{k_y}{\gamma(k_x\beta-\omega)}=\frac{\sin\theta\sqrt{1-\beta^2}}{1-\beta\cos\theta}.
\end{equation} 
For $\beta=\cos\theta$, the wave has no x-component, $k_x^\prime=0$. In this case $\beta^\prime=1$. 

The general form of the EM field tensor is
\begin{equation}
F=\left(
\begin{tabular}{cccc}
$0$ & $E_x$ & $E_y$ & $E_z$ \\
$-E_x$ & $0$ & $-H_z$ & $H_y$ \\
$-E_y$ & $H_z$ & $0$ & $-H_x$ \\
$-E_z$ & $-H_y$ & $H_x$ & $0$ 
\end{tabular}
\right).
\end{equation}
For an S-polarized EM wave and $\beta^\prime$ given by (\ref{beta^prime}) 
the EM field tensor after two Lorentz transformations ($B_x$ and $B_y$) takes the following form:
\begin{equation}
F^{\prime\prime}=\left(
\begin{tabular}{cccc}
$0$ & $0$ & $0$ & $E_0\gamma(\cos\theta-\beta)$ \\
$0$ & $0$ & $0$ & $-E_0\gamma(\cos\theta-\beta)$ \\
$0$ & $0$ & $0$ & $0$ \\
$-E_0\gamma(\cos\theta-\beta)$ & $E_0\gamma(\cos\theta-\beta)$ & $0$ & $0$ 
\end{tabular}
\right).
\end{equation}
Note that, for $\beta=\cos\theta$, the field vanishes, corresponding to $\beta^\prime=1$. 

The equation of motion for the foil, assuming $|\rho|=1$, i.e. total reflection, 
and taking into account the fact that $n_e^{\prime\prime}l^{\prime\prime}=n_e l/\cos\theta^\prime$, 
which is due to the second Lorentz boost ($B_y(\beta^\prime)$), is
 
\begin{equation}\label{eq of motion 1}
\frac{d\beta}{dt}=\kappa\frac{\sqrt{1-\beta^2}(\cos\theta-\beta)^3}{1-\beta\cos\theta}.
\end{equation}
We notice that the r.h.s. of (\ref{eq of motion 1}) is proportional to $(\cos\theta-\beta)$, \textit{i.e.}, the radiation pressure force acting on the foil vanishes for $\cos\theta=\beta$ and there is a fundamental limit on the maximum attainable ion energy arising from the angle of incidence being larger than zero. The equation of motion for the foil under the action of the oblique incident EM wave with $\beta_g<1$ is

\begin{equation}\label{eq of motion 3}
\frac{d\beta}{dt}=\kappa\frac{\sqrt{1-\beta^2}(\beta_g\cos\theta-\beta)[(\beta_g\cos\theta-\beta)^2+(1-\beta^2)(1-\beta_g^2)]}{1-\beta\beta_g\cos\theta}.
\end{equation}
The r.h.s. of this equation is proportional to $(\beta_g\cos\theta-\beta)$, \textit{i.e.}, $\beta_{max}=\beta_g\cos\theta$. We see that the two limitations 
(group velocity and angle of incidence) are combined in a single limit for the maximum ion energy. 
In Fig. \ref{fig-5} we present the results of the numerical solution of the equation of motion in two cases: (i) RPA 
and (ii) RPA with both oblique incidence and finite group velocity. These two curves demonstrate a behavior similar to the one described in section 3: 
the PRA case leads to unconstrained energy growth, while the RPA with both oblique incidence and finite group velocity shows the existence of the maximum attainable ion energy. 

In order to characterize the asymptotic behavior of the foil velocity Eq. (\ref{eq of motion 1}) is solved in quadratures:
\[
\frac{1}{2}\left\{\frac{\sqrt{1-\beta^2}}{(\cos\theta-\beta)^2}+\frac{\sqrt{1-\beta^2}\tan^{-1}\theta/\sin\theta}{(\beta-\cos\theta)}-\frac{4\cos 2\theta}{\sin^2 2\theta}\right. \]
\begin{equation}
\left.+\frac{1}{\sin^3\theta}\log\left[\frac{\cos\theta(1-\beta\cos\theta+\sqrt{1-\beta^2}\sin\theta)}{(\cos\theta-\beta)(1+\sin\theta)}\right]\right\}
\end{equation}
\[=K_\beta(t),\]
If we assume that $a(t)=$constant, then for $t\rightarrow\infty$, we obtain the scaling for the foil velocity
\begin{equation}
\beta=\cos\theta-\sqrt{\frac{\sin\theta}{2\kappa t}}.
\end{equation}
Thus the solution of Eq. (\ref{eq of motion 1}) in the limit $t\rightarrow\infty$ demonstrate a power law dependence as $\beta\rightarrow\beta_{max}$.

Equation (\ref{eq of motion 3}) can also be solved in quadratures:
\[
\frac{1}{(1-\beta_g^2)\sqrt{1-\beta_g^2\cos^2\theta}}\left\{-\log\left[\frac{\beta_g\cos\theta-\beta}{\beta_g\cos\theta}\right]\right.
\]
\begin{equation}
\left .+\log\left[\frac{1-\beta\beta_g\cos\theta+\sqrt{\beta_g^2\cos^2\theta}\sqrt{1-\beta^2}}{1+\sqrt{\beta_g^2\cos^2\theta}}\right]\right\}
\end{equation}
\[
+\frac{\beta_g^2}{1-\beta_g^2}\left\{\frac{1}{A}\log\frac{(\beta_g^2+C\beta+A\sqrt{1-\beta^2})C}{(C+\beta_g\beta)(\beta_g^2+A)}\right\}=K_\beta(t).
\]
Here
\begin{equation}
A=\sqrt{1-\beta_g^2}-i\beta_g^2\cos\theta\sin\theta~~~\text{and}~~~C=-\beta_g-i\sqrt{1-\beta_g^2}\beta_g\sin\theta.
\end{equation}
If we assume $a(t)=$constant, then for $t\rightarrow\infty$
\begin{equation}
\beta=\beta_g\cos\theta\left\{1-\exp\left[-(1-\beta_g^2)(1-\beta_g^2\cos^2\theta)^{1/2}\kappa t\right]\right\}.
\end{equation}
The maximum velocity in this case is determined by the combination of the laser pulse group velocity and the angle of incidence, $\beta_{max}=\beta_g\cos\theta$. The solution of Eq. (\ref{eq of motion 3}) in the limit $t\rightarrow\infty$ demonstrate the exponential dependence as the value of the foil velocity approaches its maximum.  

\begin{figure*}[h!]
\centering
\epsfxsize15cm\epsffile{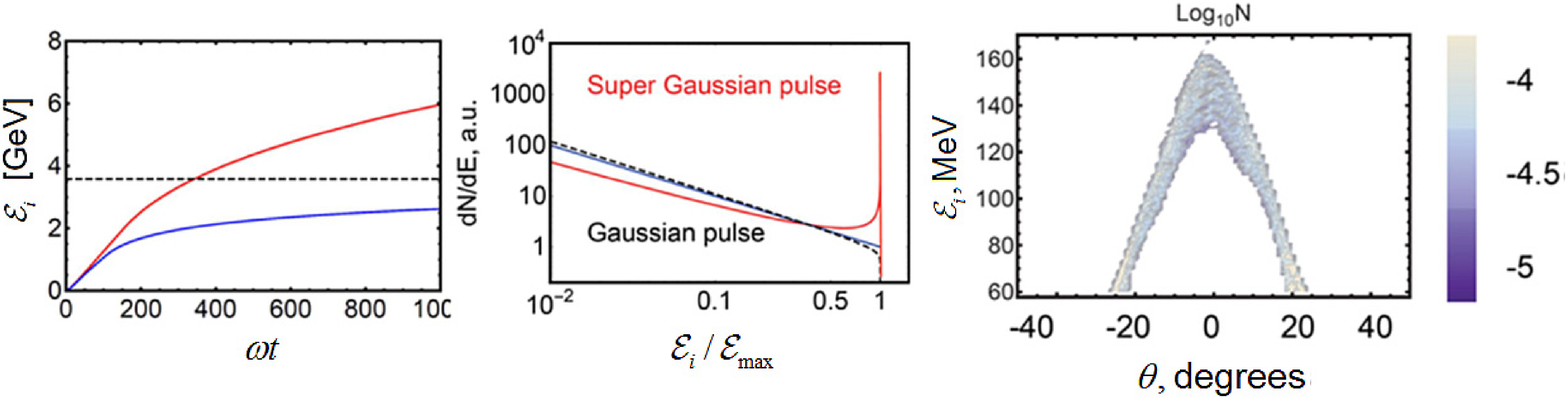}
\caption{\label{fig-5} (a) The dependence of the maximum ion energy on time for $\beta_g=1$ (red curve) and $\beta_g<1$ (blue curve) 
for a 1.8 PW laser pulse, angle of incidence $\theta=\pi/16$, and group velocity $\beta_g=0.995$. 
The foil is composed of solid density $n_e=400 n_{cr}$ hydrogen and has a thickness of $0.25\lambda$. 
(b) The characteristic spectra of protons for Gaussian and super-Gaussian pulses. 
(c) The distribution of the number of protons in the (angle, energy) plane for a 13 J laser pulse interacting with a $n_e=400 n_{cr}$ hydrogen foil, obtained in Particle-in-Cell simulations, using code REMP \cite{REMP}.}
\label{fig-5}
\end{figure*}

The off-normal incidence equation of motion can be applied for the characterization of the RPA of deformed targets. In this case the target can be considered as a collection of unit surface elements each of those having its own angle between the target normal and local EM wave vector. Such picture can be used for the self-consistent description of the target deformation and expansion during the RPA. In order to show the necessity of such approach, let us consider a naive model of the proton energy distribution. We assume that the energy of protons depends on transverse coordinate the same way as the laser pulse intensity, 
${\cal E}={\cal E}_{max}\exp\left(-2 r^\alpha/w^\alpha\right)$, where ${\cal E}_{max}$ is the maximum proton energy, which is achieved on axis. Here $\alpha=2$ stands for a Gaussian pulse and $\alpha=4,6,8...$ stands for a super-Gaussian pulse. Using this distribution and assuming a homogenous distribution of protons in the transverse direction, we can plot the spectra of protons in the Gaussian and super-Gaussian cases (see Fig. 5b). This model results in the prediction that the super-Gaussian pulses should produce proton beams with the monoenergetic feature at high energies, while the Gaussian pulses lead to the exponentially decaying spectra of protons. However, if we compare this prediction to the results of computer simulations (see Fig. 5c) of a super-Gaussian pulse interaction with a thin foil, we see no enhancement of the number of protons at high energies. Moreover, the transverse distribution of proton energy does not follow the super-Gaussian profile of the laser pulse intensity. This observation is likely due to the deformation of the target in the course of the interaction and the resulting effects of off-normal incidence. Further investigation of such interaction will be reported elsewhere.

\section{Conclusions} 
In this paper we studied the factors that limit the effectiveness of radiation pressure acceleration. The RPA regime is very attractive for applications due to its high effectiveness, however this effectiveness can be reduced by a number of factors. We identified (i) target transparency that leads to a reduced coupling of the laser pulse to the target and thus reduced maximum ion energy, (ii) finite group velocity, which imposes a fundamental limit on the maximum attainable ion energy, (iii) transverse expansion, which leads to increased target transparency and thus reduced maximum ion energy, and (iv) off-normal incidence, which imposes limit on maximum attainable ion energy, determined by the angle of incidence. Though some of these factors lead to severe limits on the maximum attainable ion energy, we showed that laser pulse tailoring, or special target design, may compensate these limits, enabling the laser acceleration of high-energy ions.  

The target transparency plays a crucial role in the ion acceleration, especially in the case of RPA. If the target is transparent for radiation, then most of the laser pulse is transmitted through the target without affecting it. In the opposite case of the target being opaque, this opaqueness is maintained by a large number of ions. Thus the energy gain of an individual ion, which is equal to laser pulse energy divided by the number of accelerated ions in the most favorable case, is suppressed. It is well established that the ion acceleration is most effective at $a_0=\varepsilon_p$, \textit{i.e.}, at the threshold of target transparency. This relation is easily understood. The pulse can push the foil without going through, and this opacity is maintained by a minimum possible number of ions, increasing the individual ion energy gain. We established the relation between $a_0$ and $\varepsilon_p$, taking into account relativistic effects, $a_0=\gamma\varepsilon_p$, which gives guidance to what laser-target parameters should be used to maximize the ion energy gain. This condition also enables one to optimize the ion acceleration by reducing the acceleration length. Optimization is achieved by choosing the laser pulse profile in such a way that this condition is satisfied at every instant in time during the acceleration process. We illustrated the fact that such a laser profile can be obtained both analytically and numerically.  

The laser pulse group velocity  being smaller than the vacuum light speed, imposes a fundamental limit on the maximum attainable ion velocity (less than the laser group velocity). This limit is connected with the fact that in the rest frame of the target moving with the velocity equal to the group velocity of the EM wave the wavevector vanishes and so does the radiation pressure. The group velocity effects will manifest themselves in the case of tightly focused laser pulses, which are usually employed to reach the RPA regime. Also if the laser pulse prior to interacting with a foil, has to penetrate through the pre-plasma, created by the pre-pulse, it would affect the value of its group velocity. Several schemes of RPA realization involve guiding the laser pulse through some pre-formed channel to the target. In this case the group velocity of the laser will be determined by the properties of the guiding structure. Though the current level of laser accelerated ion energies is too low to observe the effect of group velocity, the next generation laser facilities, delivering pulses of PW and multi-PW levels and able to generate hundreds MeV or even GeV ion beams, will be able to probe the group velocity dependence of the maximum ion energy. 

Above the RPA regime is usually realized with tightly focused laser pulses to reach high intensity and ultra-thin foils to maximize energy gain per ion. However in the case of such laser pulse interaction with such foils, the finite spot size effects begin to manifest themselves limiting the maximum attainable ion energy. These effects are due to the deformation of the foil caused by the transverse intensity profile of a laser pulse, and It results in the stretching of the irradiated spot as it is being pushed by the pulse. Consequently the area density of this spot decreases to the point when it becomes transparent to the laser. At this point the acceleration is effectively terminated. We showed that typically the transverse expansion dominates over the group velocity, severely limiting the maximum attainable ion energy. Of course the point of termination depends on the radius of the laser focal spot. The larger is the radius the smaller is the transverse expansion. However the maximum ion energy depends on the laser intensity - the larger is the intensity, the higher is the ion energy. Thus for a given laser energy and foil there should be an optimal focal spot radius that maximizes ion energy. Apart from this optimization it is possible to enhance the accelerated ion energy by modifying the target. The modification should preserve the surface density of the foil throughout the acceleration process. We proposed to employ an external guiding, which is realized through the utilization of a composite target consisting of a foil with a near critical density (NCD) slab attached to the foil back side. In the case of such a target the laser will accelerate an irradiated spot of the foil through the NCD plasma. While propagating in the NCD plasma the pulse will generate a channel in electron and ion density, analogous to the case of the MVA regime of ion acceleration. Though the foil density will drop due to the transverse expansion, the NCD plasma electrons being snowplowed by the pulse would provide an opaque density spike that, being pushed by the radiation pressure, drag the ions of the foil with it. Thus the laser pulse will be guided without the intensity loss due to diffraction, and the accelerated foil will stay opaque. The final ion energy will be determined by the group velocity of the laser in the NCD plasma. 

In many cases the laser ion acceleration experiments are carried out with the laser incident on a foil at some angle to the normal to the target surface. It is usually done to avoid the backreflection of the laser light into the system and consequent damage of it. Some mechanisms of ion acceleration demonstrate the dependence of the acceleration process on the angle of incidence, and this motivates the study of off-normal incidence effects in the case of the RPA. We found that the maximum attainable ion velocity is equal to $\beta_g\cos\theta$. Thus the off-normal incidence modifies the group velocity limit. As one can see the effect of the off-normal incidence in the case of the RPA is quite strong and should manifest itself in future experiments at PW-class laser facilities. 

Our analytical results on the off-normal incidence in the RPA regime can also be used to characterize the acceleration of a foil being deformed by the radiation pressure. We briefly discussed that, as acceleration evolves, the foil transforms from a flat surface to a curved surface surrounding the laser pulse. Then the radiation pressure exerted on each surface element can be characterized by its own angle of incidence, \textit{i.e.}, the angle between local normal to the surface and local direction of the wavevector. Thus the farther is the surface element from the laser central axis the more the radiation pressure is suppressed by the factor $\cos\theta$. We used this to explain the discrepancy between a simple analytical model, which predicts the ion spectrum assuming the ion energy distribution follows the laser intensity transverse profile, and the results of PIC simulations. Whereas the simple model predicts a peak in the high energy part of the spectrum, there is none to be found in the PIC results. It is due to the suppression of the radiation pressure by the cosine of the local angle of incidence.

\begin{figure}[h!]
\centering
\epsfxsize10cm\epsffile{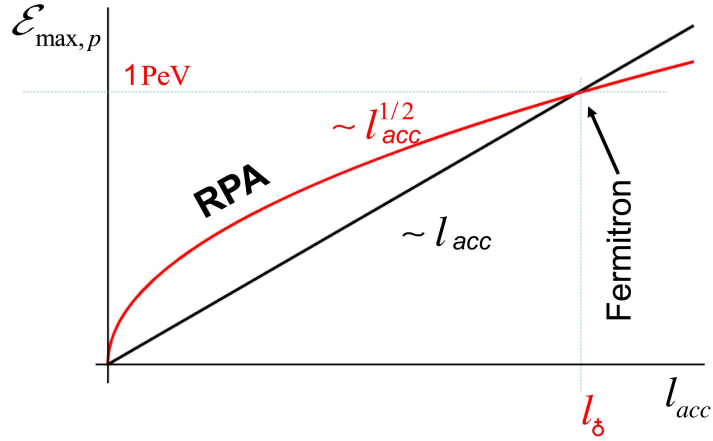}
\caption{ Upper energy limits for standard and laser ion accelerators.}
\label{FERMIvsVEKSLER}
\end{figure}

In conclusion we make a remark of general character concerning the maximum-maximorum ion energy achievable in a terrestrial environment.  It is well known, each new generation of accelerators aimed at basic research is characterized by parameters greatly exceeding the ones of the previous generation. However, ultimately, there are questions of the accelerator size and cost. Regarding the question of the maximum achievable size (and the
maximum energy of charged particles), Enrico Fermi pointed out in his speech ``What can we learn with high energy accelerators?'' delivered on January 29, 1954 that the maximum energy would be approximately $10^{15}$ eV for an accelerator length equal to the circumference of Earth's equator (40,075 km), which is equal to the maximal accelerator size on Earth \cite{JWC}. This follows from the requirement that the amplitude of the accelerating electric
field in a standard accelerator $\approx 100$ MeV/m does not cause a vacuum breakdown. Such the hypothetical accelerator is called the ``Fermitron'' or ``Globatron'', although E. Fermi named his concept as an ``Ultimate Accelerator''. In the case of the RPA regime, the ion final energy ${\cal E}_{ion}=m_i c^2\gamma_{ion}$ and the laser pulse length $l_{las}=N_C \lambda$ (here we have expressed the laser length via the laser wavelength and the number of cycles in the pulse, $N_c$) are related to the acceleration length $l_{acc}$ as
\begin{equation}
l_{acc}=2 \gamma_{ion}^2 l_{las},
\end{equation}
then $\gamma_{ion}\sim\sqrt{l_{acc}/N_C\lambda}$. For proton accelerator of the size about $l_{acc}=4\times10^9$ cm, for the laser pulse length $\lambda N_C=10\times 10^{-4}$ cm, where $\lambda=1$ $\mu$m, the maximum achievable energy is $5\times10^{15}$ eV, which is the same order of magnitude 
as the energy gain in the Fermitron, the conventional accelerator going around the Earth equator. This is due to the fact that the ion energy provided by (linear) conventional accelerators is linearly proportional to the accelerator length, while the ion energy from laser accelerators scales as a square root of the accelerator length, illustrated in Fig. \ref{FERMIvsVEKSLER}. For the accelerator length smaller than the Earth equator circumference the laser acceleration has a potential to provide higher energy of accelerated ions. There is an apparent way to achieve larger RPA accelerated ions by shortening the laser pulse length via decreasing the number of cycles per pulse or/and decreasing the wavelength of the electromagnetic radiation.

\section*{Acknowledgements}

We acknowledge support of the Director, Office of Science, office of High Energy Physcis, of the US DOE under Contract No. DE-AC02-05CH11231 and the Ministry of Education, Youth and Sports of the Czech Republic (ELI-Beamlines reg. No. CZ.1.05/1.1.00/02.0061). The authors would like to thank for discussions C. Benedetti, M. Chen, C. G. R. Geddes, Q. Ji, S. Steinke, L. Yu, D. Margarone, and G. Korn.

\end{document}